\documentclass[aps,pre,twocolumn,groupedaddress]{revtex4}

\usepackage{graphicx}
\usepackage{dcolumn}
\usepackage{bm}
\usepackage{subfigure}

\begin{document}


\title{Beyond the Imry-Ma Length: Scaling Behavior in the 3D Random Field $XY$ Model}


\author{Ronald Fisch}
\email[]{ronf124@gmail.com}
\affiliation{382 Willowbrook Dr.\\
North Brunswick, NJ 08902}


\date{\today}

\begin{abstract}
We have performed studies of the 3D random field $XY$ model on
$L \times L \times L$ simple cubic lattices with periodic boundary conditions,
with a random field strength of $h_r$ = 1.875, for $L =$ 64, 96 and 128,
using a parallelized Monte Carlo algorithm.  We present results for the
angle-averaged magnetic structure factor, $S ( k )$ at $T$ = 1.00, which
appears to be the temperature at which small jumps in the magnetization per
spin and the energy per spin occur.  The magnetization jump per spin scales
with size roughly as $L^{- 3/4}$, while the energy jump per spin scales like
$L^{- 3/2}$.  The results also indicate the existence of an approximately
logarithmic divergence of $S ( k )$ as $k \to 0$.  The magnetic
susceptibility, $\chi (\vec{\bf k} = 0 )$, on the other hand, seems to have
a value of about 14.2 under these conditions.  This suggests the absence of
a ferromagnetic phase, and that the lower critical dimension for long-range
order in this model is three.  Similar results are found for $L$ = 64 samples
at $h_r$ = 2.0 and $T$ = 0.875.  We expect that the behavior is qualitatively
similar along the entire phase boundary, but the scaling exponents may not be
universal.  These results appear to be related to recent work on quantum disorder.

\end{abstract}


\maketitle

\section{Introduction}

The behavior of the three-dimensional (3D) random-field $XY$ model
(RFXYM) at low temperatures and weak to moderate random field
strengths continues to be controversial.  A detailed calculation by
Larkin\cite{Lar70} showed that, in the limit that the number of
spin components, $n$, becomes infinite, the ferromagnetic phase
becomes unstable when the spatial dimension of the lattice is less
than or equal to four, $d \le 4$.  Dimensional reduction
arguments\cite{IM75,AIM76} appeared to show that the long-range order
is unstable for $d \le 4$ for any finite $n \ge 2$.  However, there
are several reasons for questioning whether dimensional reduction
can be trusted for $XY$, {\it i.e.} $n = 2$, spins.

Some time ago, Monte Carlo calculations\cite{GH96,Fis97} showed
that there was a line in the temperature vs. random-field plane
of the phase diagram of the three-dimensional (3D) random-field
$XY$ model (RFXYM), at which the magnetic structure factor becomes
large as the wave-number $k$ becomes small.  Additional calculations\cite{Fis07}
indicated that there appeared to be small jumps in the magnetization
and the energy of $L$ = 64 lattices at a random field strength of
$h_r = 2.0$, at a temperature somewhat below $T = 1.0$.  Further
calculations\cite{Fis10} showing similar behavior for other values
of the random field strength were also performed. If such behavior
persisted for larger values of $L$, with the sizes of these jumps
being independent of $L$ for large $L$, this would demonstrate that
there is a ferromagnetic phase at weak to moderate random fields and
low temperatures for this model.  However, Aizenman and Wehr\cite{AW89,AW90}
have proven that this cannot happen in 3D.  The sizes of these jumps
must scale to zero as $L$ goes to infinity.  The rates of the scaling
characterizes the phase transition, analogous to the critical exponents
which describe critical behavior in second order phase transitions.

Since there have been substantial improvements in computing hardware
and software over the last ten years, the author felt it worthwhile
to conduct a new Monte Carlo study of this model using parallel
processing.  The results of that study for $L \times L \times L$
simple cubic lattices with $L =$ 64, 96 and 128 will be presented
here.  Extending the results to larger sizes does not appear to be
practical at this time.

\section{The Model}

For fixed-length classical spins the Hamiltonian of the RFXYM is
\begin{equation}
  H ~=~ - J \sum_{\langle ij \rangle} \cos ( \phi_{i} - \phi_{j} )
  ~-~ h_r \sum_{i} \cos ( \phi_{i} - \theta_{i} )  \, .
\end{equation}
Each $\phi_{i}$ is a dynamical variable which takes on values
between 0 and $2 \pi$. The $\langle ij \rangle$ indicates here a sum
over nearest neighbors on a simple cubic lattice of size $L \times L
\times L$. We choose each $\theta_{i}$ to be an independent
identically distributed quenched random variable, with the
probability distribution
\begin{equation}
  P ( \theta_i ) ~=~ 1 / 2 \pi   \,
\end{equation}
for $\theta_i$ between 0 and $2 \pi$.  We set the exchange constant to
$J = 1$.  This gives no loss of generality, since it merely fixes the
temperature scale.  This Hamiltonian is closely related to models of
vortex lattices and charge density waves.\cite{GH96,Fis97}

Larkin\cite{Lar70} studied a model for a vortex lattice in a
superconductor.  His model replaces the spin-exchange term of the
Hamiltonian with a harmonic potential, so that each $\phi_{i}$ is no
longer restricted to lie in a compact interval.  He argued that for
any non-zero value of $h_r$ this model has no ferromagnetic phase on
a lattice whose dimension $d$ is less than or equal to four.  The Larkin
approximation is equivalent to a model for which the number of spin
components, $n$, is sent to infinity.  A more intuitive derivation
of this result was given by Imry and Ma,\cite{IM75} who assumed that
the increase in the energy of an $L^d$ lattice when the order parameter
is twisted at a boundary scales as $L^{d - 2}$ for all $n > 1$, just as
it would for $h_r = 0$.  Using this assumption, they argued that when
$d \le 4$ there is a length $\lambda$, now called the Imry-Ma length,
at which the energy which can be gained by aligning a spin domain with
its local random field exceeds the energy cost of forming a domain wall.
From this they claimed that the magnetization would decay to zero when
the system size, $L$, exceeds $\lambda$.

Within a perturbative $\epsilon$-expansion one finds the phenomenon
of ``dimensional reduction".\cite{AIM76} The critical exponents of any
$d$-dimensional $O(n)$ random-field model appear to be identical to
those of an ordinary $O(n)$ model of dimension $d - 2$. For the
$n = 1$ (RFIM) case, this was soon shown rigorously to be incorrect for
$d < 4$.\cite{Imb84,BK87}  More recently, extensive numerical results
for the Ising case have been obtained for $d = 4$ and
$d=5$.\cite{FMPS17a,FMPS17b}  They determined that dimensional reduction
is ruled out numerically in the Ising case for $d = 4$, but not for
$d = 5$.\cite{FMPS18}

The scaling behavior is somewhat different for $n \ge 2$.  Because
translation invariance is broken for any non-zero $h_r$, it seems quite
implausible to the current author that the twist energy for Eqn.~(1)
scales as $L^{d - 2}$ for large $L$ when $d \le 4$, even though this is
correct to all orders in perturbation theory.  The problem with assuming
this scaling is that the Irmy-Ma length provides a natural length scale
to the problem. We need to scale out to the Imry-Ma length before we can
learn the true long-distance behavior of the model.  This means that the
effective strength of the randomness cannot be assumed to grow without
bound when $d \le 4$, even though it grows for weak non-zero $h_r$.
We must do an detailed calculation to find out what actually happens.

An alternative derivation of the Imry-Ma result by Aizenman and
Wehr,\cite{AW89,AW90} which claims to be mathematically rigorous,
also makes an assumption equivalent to translation invariance.
Although the average over the probability distribution of random
fields restores translation invariance, one must take the infinite
volume limit first.  It is not correct to interchange the infinite
volume limit with the average over random fields.  This problem of
the interchange of limits is equivalent to the existence of replica
symmetry breaking.  The existence of replica symmetry breaking in
random field models was first shown by Mezard and Young,\cite{MY92}
about two years after the work of Aizenman and Wehr.  Mezard and
Young emphasized the Ising case, and the fact that this applies for
all finite $n$ seems to have been overlooked by many people for a
number of years.  A functional renormalization group calculation
going to two-loop order was performed by Tissier and Tarjus,\cite{TT06}
and independently by Le Doussal and Wiese.\cite{LW06}  They found
that there was a stable critical fixed point of the renormalization
group for some range of $d$ below four dimensions in the $n = 2$
random field case.  However, it is not clear from their calculation
what the nature of the low-temperature phase is, or whether this
fixed point is stable down to $d = 3$.  Tarjus and Tissier\cite{TT08}
later presented an improved version of this calculation, which
explains more explicitly why dimensional reduction fails for the
$n = 2$ case when $d \leq 4$.

\section{Structure factor and magnetic susceptibility}

The magnetic structure factor, $S (\vec{\bf k}) = \langle
| \vec{\bf M}(\vec{\bf k}) |^2 \rangle $, for $XY$ spins is
\begin{equation}
  S (\vec{\bf k}) ~=~  L^{-3} \sum_{ i,j } \cos ( \vec{\bf k} \cdot
  \vec{\bf r}_{ij}) \langle \cos ( \phi_{i} - \phi_{j}) \rangle  \,   ,
\end{equation}
where $\vec{\bf r}_{ij}$ is the vector on the lattice which starts
at site $i$ and ends at site $j$, and here the angle brackets denote
a thermal average.  For a random field model, unlike a random bond
model, the longitudinal part of the magnetic susceptibility, $\chi$,
which is given by
\begin{equation}
  T \chi (\vec{\bf k}) ~=~ 1 - M^2 ~+~ L^{-3} \sum_{ i \ne j } \cos (
  \vec{\bf k}  \cdot \vec{\bf r}_{ij}) (\langle \cos ( \phi_{i} - \phi_{j}
  ) \rangle ~-~ Q_{ij} )  \,   ,
\end{equation}
is not the same as $S (\vec{\bf k})$ even above $T_c$.  For $XY$ spins,
\begin{equation}
  Q_{ij} ~=~ \langle \cos ( \phi_{i} ) \rangle \langle \cos (
  \phi_{j} ) \rangle ~+~ \langle \sin ( \phi_{i} ) \rangle \langle
  \sin ( \phi_{j} ) \rangle  \,  ,
\end{equation}
and
\begin{equation}
  M^2 ~=~ L^{-3} \sum_{i} Q_{ii}
      ~=~ L^{-3} \sum_{i} [ \langle \cos ( \phi_{i} ) \rangle^2 ~+~
       \langle \sin ( \phi_{i} ) \rangle^2 ] \,  .
\end{equation}
When there is a ferromagnetic phase transition, $S ( \vec{\bf k} = 0 )$
has a stronger divergence than $\chi ( \vec{\bf k} = 0 )$.

The scalar quantity $\langle M^2 \rangle$, when averaged over a set
of random samples of the random fields, is a well-defined function
of the lattice size $L$ for finite lattices. With high probability,
it will approach its large $L$ limit smoothly as $L$ increases.
The vector $\vec{\bf M}$, on the other hand, is not really a
well-behaved function of $L$ for an $XY$ model in a random field.
Knowing the local direction in which $\vec{\bf M}$ is pointing,
averaged over some small part of the lattice, may not give us a
strong constraint on what $\langle \vec{\bf M} \rangle$ for the
entire lattice will be.  When we look at the behavior for
all $\vec{\bf k}$, instead of merely looking at $\vec{\bf k} = 0$,
we get a much better idea of what is really happening.

\section{Numerical results for $S ( k )$ and $\chi (\vec{\bf k} = 0)$}

In this work, we will present results for the average over angles of
$S (\vec{\bf k})$, which we write as $S (k)$.  The data were
obtained from $L \times L \times L$ simple cubic lattices with $L =$
64, 96 and 128 using periodic boundary conditions.  The calculations
were done using a clock model which has 8 equally spaced dynamical
states at each site.  In addition, there is a static random phase at
each site which was chosen to be $0, \pi/12$ or $\pi/6$ with equal
probability.  Tests were also conducted for clock models which had 6
dynamical states at each site, and these were found not to be good
enough quantitative approximations to the limit of a large number of
dynamical states per site.

The idea of adding $p$-fold symmetry-breaking terms to an $XY$ model
goes back to Jose, Kadanoff, Kirkpatrick and Nelson,\cite{JKKN77} who
studied the effects of nonrandom fields of this type on the
Kosterlitz-Thouless (KT) transition in 2D.  The result they found was
that the KT transition survives the addition of terms of this type near
$T_c$ if $p > 4$, but that the system becomes ferromagnetic at some
lower value of $T$.  This work was extended to $p$-fold fields which
varied randomly in space by Houghton, Kenway and Ying\cite{HKY81} and
Cardy and Ostland.\cite{CO82}  It was found that the KT transition
survives in the random $p$-fold field case for $p \ge 3$.

Generalizing this idea to $d > 2$ is straightforward.  It has been
known for some time that a nonrandom $Z_p$ model of this type is in the
universality class of the ferromagnetic $XY$ model whenever $p > 4$.\cite{WK74}
For random-phase $Z_p$ models without a random-field term, there are no
analytical results.  However, it has been found numerically that in 3D the
model is in the universality class of the pure $XY$ model under most conditions,
even if the number of dynamical states of each spin is only 3.\cite{Fis92}
Under conditions of very low temperature, this model may undergo an
incommensurate-to-commensurate type of charge-density wave phase transition.
Thus it is expected that, when we include the random-field term, the model
will behave essentially as a random-field $XY$ model, as long as we do not
attempt to work at very low temperatures and random field strengths much
weaker than the ones used here.\cite{Fis97}  However, we want to have more
than merely being in the same universality class, which only requires 3
dynamical states at each site.  We have found that if we use at least 8
dynamical states at each site, then the results we find numerically do not
depend on the number of dynamical states, at least for $T \ge 1.00$.

Based on earlier Monte Carlo calculations,\cite{GH96,Fis07} we know the
approximate location of the phase boundary in the ($h_r , T$) plane.
This is true despite the fact that we are not certain what the nature of
the low temperature phase is.  The reason why this is possible is that
we are able to locate the phase boundary by finding where the static
ferromagnetic correlation length first diverges as we lower $T$ or $h_r$.
It was not known {\it a priori} if it would be possible to do calculations
under conditions where we could get past the crossover region and see the
large lattice behavior on the phase boundary.

The strength of the random field for which data were obtained initially
was chosen to be $h_r$ = 1.875.  This value was picked in order to make
the value of $T_c$ close to 1.00.  This is about as low a $T_c$ as can be
used to study $L$ = 128 lattices near $T_c$ with the computing resources
available, since relaxation times increase as $T_c$ decreases.

The direction of the random field at site $i$, $\theta_i$, was chosen
randomly from the set of the 24th roots of unity, independently at each
site.  Since $\theta_i$ has 24 possible values, our past experience with
models of this type indicates that there is no reason to expect that the
discretization will affect the behavior near $T$ = 1.00 in an observable
way.  Later, the program was modified so that $\theta_i$ had 48 possible
values, and each $\phi_i$ had 12 allowed equally spaced states.  Naturally,
this modification caused the program to run more slowly, but the changes
in the numerical results caused by using a finer discretization for the
same sequence of $\theta_i$ were small.  Checks like these have been done
before by this author and a number of other authors on various $XY$ models,
so this was completely expected.  As discussed later, the modified program
was also used to obtain data for $L$= 64 lattices with $h_r$ = 2.00 at
$T = 0.875$.  It is necessary to use a finer mesh as the temperature is
lowered, in order for the approximation to an $XY$ model to remain
quantitatively accurate.

The computer program uses three independent pseudorandom number generators:
one for choosing initial values of the dynamical variables, $\phi_i$, in
the hot start initial condition, one for setting the static random phases,
$\theta_i$, and a third one for the Monte Carlo spin flips, which are
performed by a single-spin-flip heat-bath algorithm.

The pseudorandom number generators for the $\phi_i$ and the $\theta_i$ are
standard linear congruential generators which have been used for many years.
Given the same initial seeds, they will always produce the same string of
numbers, which is a property needed by the program.  They have excellent
statistical properties for strings of numbers up to length $10^8$ or so,
which is adequate for our purpose here.  Using separate generators for
choosing the initial values of the dynamical $\phi_i$ and the static random
$\theta_i$ was not really necessary, since the hot starts were always done
at a high value of $T$.  However, the cost of doing this is negligible, and
it would have allowed the use of random initial start conditions at any value
of $T$, although that was not done in the work reported here.

The pseudorandom number generator used for the Monte Carlo spin flips
was the library function $random\_number$ supplied by the Intel
Fortran compiler, which is suitable for parallel computation.  It is
believed that this generator has good statistical properties for
strings of length $10^{14}$, which is what we need here.  However,
the author has no ability to check this for himself.  The spin-flip
subroutine was parallelized using OpenMP, by taking advantage of the
fact that the simple cubic lattice is two-colorable.  It was run on
Intel multicore processors of the Bridges Regular Memory machine at
the Pittsburgh Supercomputer Center.  The code was checked by setting
$h_r = 0$, and seeing that the known behavior of the pure ferromagnetic
3D XY model was reproduced correctly.  It was found, however, that
using more than two cores in parallel did not result in any additional
speedup of the calculation.  This made it impractical to study 3D
lattices larger than $L = 128$.

24 different realizations of the random fields $\theta_i$ were studied for
each value of $L$.  Each lattice was started off in a random spin state at
$T = 2.25$, slightly above the $T_c$ for the pure $O(2)$ model, which is
approximately 2.202.\cite{Jan90}  The $T_c$ for a pure $Z_4$ model is 2.2557,
half that of the pure Ising model.  As far as the author knows, there are no
highly accurate calculations of $T_c$ for pure $Z_p$ models with $p > 4$ on
a simple cubic lattice.  It is expected, however, that these will converge
to the $T_c$ for the $O(2)$ model exponentially fast in $n$.  The reason for
this is that $\cos (\theta_j - \theta_i)$ for nearest neighbor $i$ and $j$
at $T_c$, which is the energy per bond at $T_c$, is 0.33 on this lattice.
This means that the typical angle between nearest neighbor spins at $T_c$ is
slightly less than $2 \pi / 5$.  Once the mesh size for $\theta_i$ becomes
less than the typical value of $\theta_j - \theta_i$, the effect of the
discretization rapidly disappears.

Each lattice was then cooled slowly to $T = 1.00$, using a cooling schedule
which depended on $L$.  Although the relaxation of the spins is not a simple
exponential function, it is quite apparent that the relaxation is becoming
very slow as $T = 1.00$ is approached.  At $T = 1.00$, the sample was relaxed
until an apparent equilibrium was reached over an appropriate time scale.
For $L$ = 64 this time scale was at least 163,840 Monte Carlo steps per spin
(MCS).  For $L$ = 96 this was increased to at least 655,360 MCS, and for
$L$ = 128 the minimum time was 1,310,720 MCS.  Some samples required
relaxation for up to three times longer than these minimum times.

The most significant fact about these times is that they are increasing
significantly with $L$.  The second important fact is the relaxation time for
these large finite sizes increases rapidly as the temperature is lowered for
$T$ near 1.00.  This indicates that what we are seeing is a cooperative effect,
similar to the critical slowing down seen at critical points.  The dynamical
relaxation behavior seen at an ordinary first-order phase transition, on the
other hand, stops slowing down once the sample size becomes larger than the
size of a nucleation droplet.

After each sample was relaxed at $T$ = 1.00, a sequence of 8 equilibrated
spin states obtained at intervals of 20,480 MCS for $L$ = 64 or 40,960 MCS
for $L =$ 96 and 128 was Fourier transformed to calculate $S (\vec{\bf k})$,
and then averaged over the sequence of 8 spin states. The data were then
binned according to the value of $k^2$, to give the angle-averaged $S ( k )$.
Finally, an average over the 24 samples was performed for each $L$.  The
average magnetization per spin at $T$ = 1.00 of these slowly cooled samples
was $0.139 \pm 0.011$ for $L$ = 64, $0.080 \pm 0.011$ for $L$ = 96, and
$0.053 \pm 0.005$ for $L$ = 128.

Data were also obtained for the same sets of samples using ordered
initial states and warming to $T$ = 1.00.  At least two, and sometimes
more initial ordered states were used for each sample.  The initial
magnetization directions used were chosen to be close to the direction
of the magnetization of the slowly cooled sample with the same set of
random fields.  This type of initial state was chosen because it was
found in the earlier work\cite{Fis07} that this is the way to find the
lowest energy minima in the phase space.  The data from the initial
condition which gave the lowest average energy for a given sample was
then selected for further analysis and comparison with the slowly cooled
state data for that sample.  The relaxation procedure at $T$ = 1.00
for the warmed states was the same one used for the cooled states, and
the calculation of $S (k)$ proceeded in the same way.  The average
magnetization per spin of these selected warmed states was $0.239 \pm 0.010$
for $L$ = 64, $0.157 \pm 0.011$ for $L$ = 96, and $0.113 \pm 0.007$ for $L$ = 128.

\begin{figure}
\includegraphics[width=3.1in]{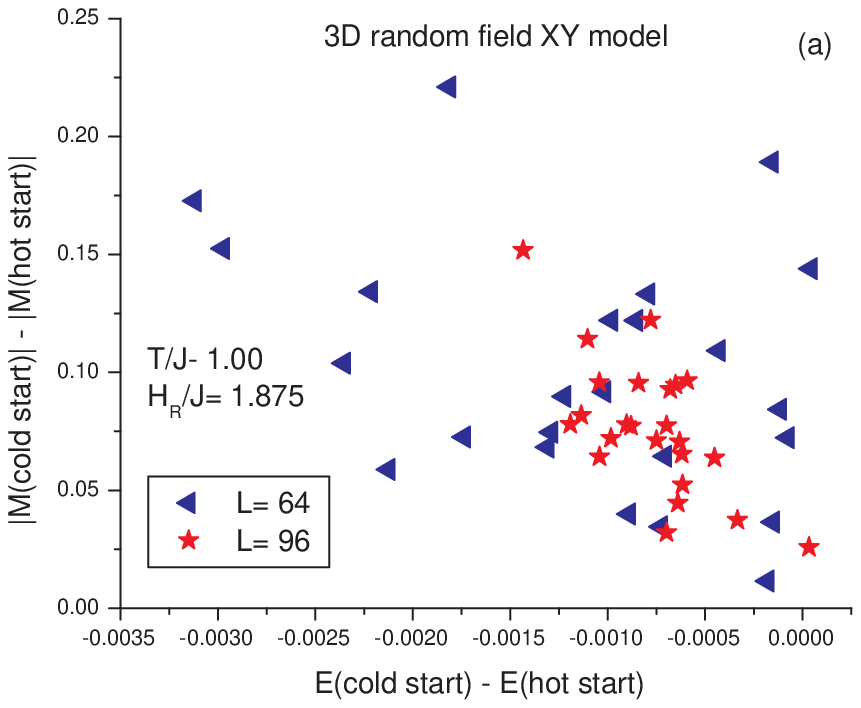}\quad
\includegraphics[width=3.1in]{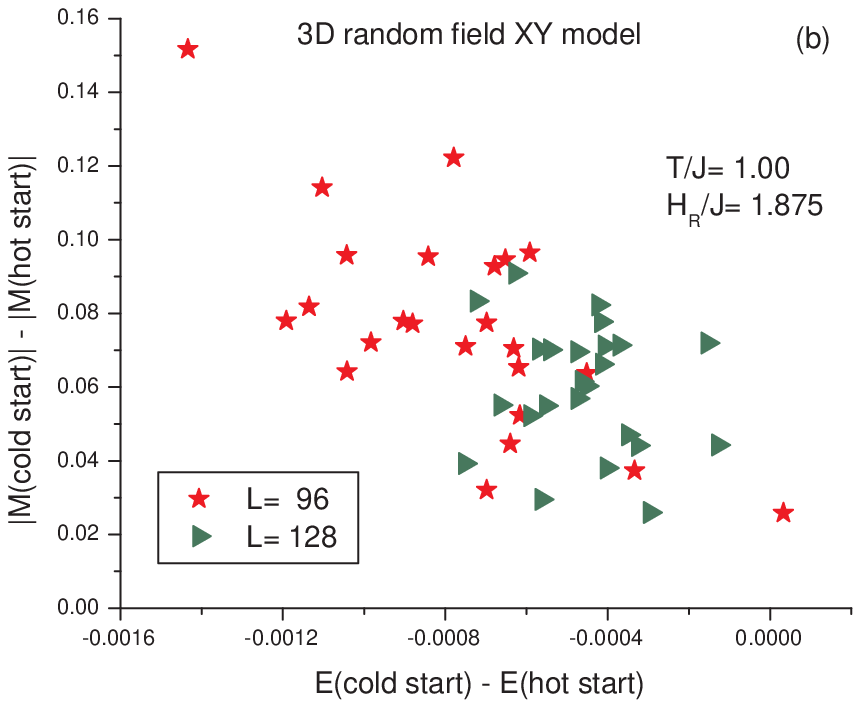}
\caption{\label{Fig.1}(color online) Jump in the magnetization vs.
jump in the energy for simple cubic lattices with $h_r$ = 1.875 at
$T$ = 1.00. (a) $L$ = 64 and $L$ = 96. (b) $L$ = 96 and $L$ = 128.
States with hot start and ordered start initial conditions are
compared for each sample.}
\end{figure}

In order to better compare the data for different sizes, the energy
per spin difference and the magnetization per spin difference between the
cooled state and the warmed state at $T$ = 1.00 were computed for each
sample.  The resulting distributions are shown in a scatter plot in Fig.~1.
Fig.~1a compares $L$ = 64 with $L$ = 96, and Fig.~1b compares $L$ = 96 with
$L$ = 128.  We see that the distributions do not show any significant
correlation between the energy difference and the magnetization difference
for the $L$ = 64 and $L$ = 128 samples.  For the $L$ = 96 samples there is a
weak tendency for the size of the jump in the magnetization to be correlated
with the size of the jump in the energy. The distribution is rather broad for
$L$ = 64, and it becomes progressively narrower as $L$ increases.

The center of the $L$ = 64 distribution is at $\delta |M| = 0.100 \pm 0.011$
and $\delta E = -0.00113 \pm 0.00019$. The center of the $L$ = 96
distribution is at $\delta |M| = 0.0773 \pm 0.0059$ and $\delta E = -0.00078
\pm 0.00006$.  For $L$ = 128 the center is at $\delta |M| = 0.0598 \pm 0.0035$
and $\delta E = -0.00046 \pm 0.00003$  The conjecture that $\delta |M|$ and
$\delta E$ will scale to zero\cite{Fis07} as $L \to \infty$ is consistent with
these data.  From these data we can make estimates of how $\delta |M|$ and
$\delta E$ behave as $L$ increases.  The $\delta E$ scaling is consistent with
$L^{- 3/2}$ behavior, as predicted by the central limit theorem.  For
$\delta |M|$, the scaling is slower than this, being roughly $L^{- 3/4}$.
More data over a wider range of $L$ would be needed before good estimates of
the rates of convergence of these parameters could be made.  Merely having
more samples of the sizes we have studied here would not really help much,
because trying to extrapolate from data over only a factor of two in $L$ is
always subject to systematic errors.  However, it seems to be true that the
scaling of $\delta |M|$ is too slow to be consistent with the central limit
theorem.  This implies that the behavior we are seeing is a true phase
transition of some kind.  At the current time, we have no theory which tells
us whether these scaling exponents should vary along the phase boundary, or
whether they should be universal.

The average specific heat of the $L$ = 64 samples at $T$ = 1.00 is
$0.7797 \pm 0.0019$ for the cooled samples, and $0.7815 \pm 0.0028$
for the heated samples.  The corresponding numbers for $L$ = 96 are
$0.7802 \pm 0.0028$ and $0.7854 \pm 0.0026$.  The fact that the specific
heat of the cooled samples is lower than the specific heat of the
somewhat more magnetized heated samples is expected.  The fact that
the difference between them is very small means that there is not much
energy associated with the disappearance of the magnetic long-range
order.  The fact that the jump in the specific heat seems to be slightly
larger for $L$ = 96 than for $L$ = 64 is normal for a weakly first-order
phase transition.  The fact that the jump is so small also means that
we are not looking at a normal second order phase transition.  Note that
the energy and specific heat of a given sample in its hot start state and
its cold start state are highly correlated.  The statistical significance
of the small difference between the hot start specific heat and the cold
start specific heat is not related to the width of the energy distribution
for different samples at $T$ = 1.00.

The uncertainty in our estimate of the $T_c$, the temperature of the
phase transition, is about an order of magnitude less than the extrapolated
shift in temperature which would be needed to make the jump in energy
between the heated samples and the cooled samples disappear.  However, the
free-energy minimum of a cold-start sample actually becomes clearly
unstable at a temperature a few percent higher than $T$ = 1.00.

Now we turn to the data for the structure factor.  The average $S ( k )$
for the 24 $L$ = 128 samples at $T$ = 1.00 is shown in Fig.~2.  $S ( k )$
is computed separately for the heated sample data and the cooled sample
data, but it is difficult to see any difference between them.  These
data are very similar to the earlier data\cite{Fis07} for $h_r$ =2 at
$T$ = 0.875.  The change in the slope of the data points now occurs near
$k$ = 0.11 instead of $k$ = 0.14, but this is about what is expected from
using the somewhat lower value of $h_r$.  From this log-log plot, it is
not clear how to extrapolate the data to small $k$.  This is due to an
inflection point in the data, when it is plotted in this way.

\begin{figure}
\includegraphics[width=3.4in]{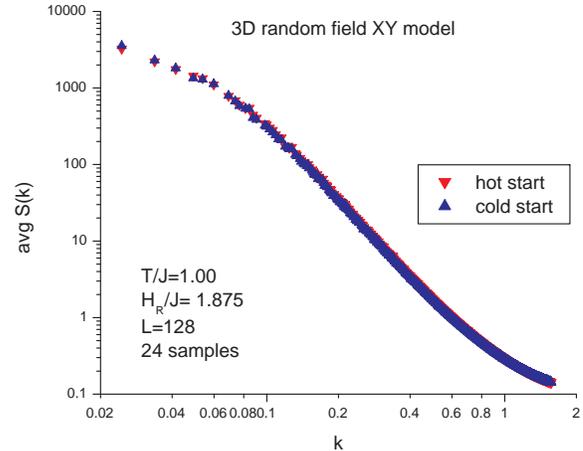}
\caption{\label{Fig.2}(color online) Angle-averaged structure factor
vs. $k$ for $128 \times 128 \times 128$ lattices with $h_r$ = 1.875 at
temperature $T$ = 1.00. Both the $x$-axis and the $y$-axis are scaled
logarithmically.  One $\sigma$ statistical errors are approximately
the size of the plotting symbols.}
\end{figure}

\begin{figure}
\includegraphics[width=3.4in]{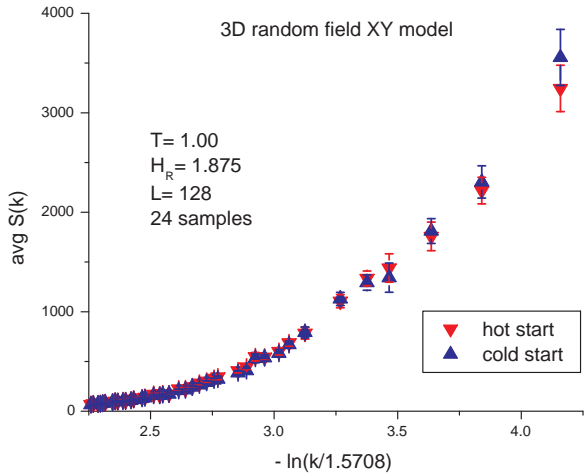}
\caption{\label{Fig.3}(color online) Angle-averaged structure factor
vs. ($-~\ln(k/1.5708)$) for $128 \times 128 \times 128$ lattices with
$h_r$ = 1.875 at temperature $T$ = 1.00.  One $\sigma$ statistical errors
are shown.}
\end{figure}

To clarify the behavior at small $k$, we replot the same $L$ = 128
data for the structure factor on a linear scale in Fig.~3.  The
scaling for the $x$-axis is chosen so that the edge of the Brillouin
zone would be at $x$ = 0, but only the small-$k$ part of the data is
shown on the graph.  When the correlation length is smaller than the
sample size, $S ( k )$ will flatten out at small $k$.  From Fig.~3
it is clear that we have no evidence for a finite correlation length
of $S ( k )$ at $L$ = 128.  However, these data also appear to rule out
the possibility that $S ( k )$ behaves like $k^{4 - \bar{\eta}}$ with
$1 < \bar{\eta} \le 2$ as $k \to 0$, which would be required for
hyperscaling to hold.  The reader who wishes to see what $S ( k )$ for
this model looks like above the phase boundary should look at the
author's 2007 paper.\cite{Fis07}

It is important to emphasize that the peak in $S ( k )$ seen for the
sample-averaged hot start data and the cold start data for these samples
is quantitatively the same, within statistical errors.  If this had not
been true, it would have meant that we had not relaxed the samples for
a long enough time.  However, it is not true that the peak looks the
same for the hot start data and the cold start data of a single sample,
before averaging.  The relaxed hot start state and relaxed cold start
state of a sample typically have a substantial overlap, but they are not
similar over the entire sample once we have reached a large enough value
of $L$.  When the size of the sample is smaller than the Imry-Ma length,
one could be misled into thinking that the system was ferromagnetic.

The author thinks it is likely that the relaxed cold start states we find
for $h_r$ = 1.875 samples at $T$ = 1.00 have a high degree of overlap with
the true ground states of these samples.  However, we have no way of
verifying this numerically.  For a sample with a small value of $h_r$, we
would not expect that the relaxed cold start state at the freezing
temperature would be as closely related to the ground state, since the
freezing transition occurs at a higher temperature.

The $L$-dependence of the magnetic susceptibility, $\chi (\vec{\bf k} = 0)$,
provides further evidence that what we are seeing does not fit the usual
scaling picture for a critical point of a random-field model, as is found in
the 3D RFIM.\cite{FMPS18}  The values of $\chi$ found using both the hot
start and cold start initial conditions at $T$ = 1.00 as a function of $L$
are shown in Table I.  It appears that $\chi$ has become almost independent
of $L$ by $L$ = 128, reaching a value of about 14.2.  According to the
universality argument of Sourlas,\cite{Sou18} this means that the phase on
the low-$T$, small-$h_r$ size of the phase boundary cannot be ferromagnetic.
It could, in principle, be true that there is another phase boundary, and a
third, low temperature phase in the phase diagram.  This is what happens at
small $h_r$ for small values of $p$.  However, the author considers this to
be implausible for the RFXYM.

\begin{quote}
Table I: Magnetic susceptibility for hot start and cold start initial
conditions at $T$ = 1.00 and $h_r$ = 1.875.
\begin{tabular}{|l|ccc|}
\hline
$L$&64&96&128\\
\hline
$\chi_{h}$&12.9 $\pm$ 0.7&14.6 $\pm$ 0.6&14.4 $\pm$ 0.7\\
$\chi_{c}$&12.6 $\pm$ 0.6&13.6 $\pm$ 0.4&14.1 $\pm$ 0.5\\
\hline
\end{tabular}
\end{quote}

A divergence of $S ( k )$ as $k \to 0$ like $\ln (k)$, or some power
of $\ln (k)$, is a strong indicator that the lower critical dimension
of the RFXYM is exactly equal to three.  The author is not aware of
other numerical results of this type of behavior in a model with
quenched random disorder, and much remains to be learned.  It would be
very exciting if similar behavior was observed by doing experiments on
physical systems which are believed to be in the universality class
of this model.

It is possible to do an explicit check to show that the discretization
is not affecting the results in a significant way.  This is done by using
a finer mesh with the same sequence of random numbers to choose the random
fields $\theta_i$.  This check was performed for two $L$ = 64 lattices, using
the 48th roots of unity.  For this check, each spin had 12 dynamical states,
and the static random phase at each site had 4 allowed values.  It was found
that the low energy minima which this program found at $T$ = 1.00 for both
the hot start and cold start initial conditions had a high degree of overlap
with the states found using the 24th roots of unity and 8 dynamical states.

The reason why the discretization works this way is that the random field
$XY$ model is not chaotic in the same way that an Edwards-Anderson spin-glass
model is\cite{FH88} when the average $J_{ij}$ is near zero.  Changing the
random field locally at a few sites, or making small, uncorrelated changes at
many sites, will not typically cause a substantial change in the low-energy
minima of the phase space.

There is nothing really special about the point $T$ = 1.00 and $h_r$ = 1.875.
The behavior anywhere along the phase boundary between the high temperature
and low temperature phases should be qualitatively the same.  This point is
merely the most convenient one for numerical calculations.  It represents the
optimal compromise between having a short crossover length ({\it i.e.} Imry-Ma
length), which happens at larger $h_r$, and a short relaxation time, which
happens at small $h_r$.  To demonstrate this explicitly, calculations at
$T$ = 0.875 and $h_r$ = 2.0 were performed for $L$ = 64.  This point was the
one used in the author's earlier work on the RFXYM.\cite{Fis07}  The earlier
work studied only 4 samples, and did not use the random static phase, but
studied a range of temperature.  Now we show data from 24 samples, using the
algorithm with 12 dynamical states and 4 values of the random static phase.

Because the relaxation along the critical line is slower at $h_r$ = 2.0 than
at $h_r$ = 1.875, the lengths of the Monte Carlo runs used for $L$ = 64 in this
case were essentially the ones used for $L$ = 96 in the $h_r$ = 1.875 case.
The sequences of random numbers used to choose the random fields at $h_r$ = 2.0
were identical to the ones used for $L$ = 64 at $h_r$ = 1.875.  Only the strength
of the fields was varied.  Because of this, there was a substantial overlap in
the states which were found in the two cases.  Making the random fields a little
stronger means that we need to go lower in $T$ to reach the phase transition,
but the nature of what is going on is mostly unchanged.  In Fig.~4 we show the
jumps in the magnetization versus the jumps in the energy.  These are similar to
the results shown in Fig.~1(a) for $L$ = 64.  The results are similar, but the
jumps are somewhat smaller at the higher value of $h_r$.  This is not surprising,
since we know that the phase transition line goes to $T$ = 0 at an $h_r$ of
roughly 2.3.

\begin{figure}
\includegraphics[width=3.1in]{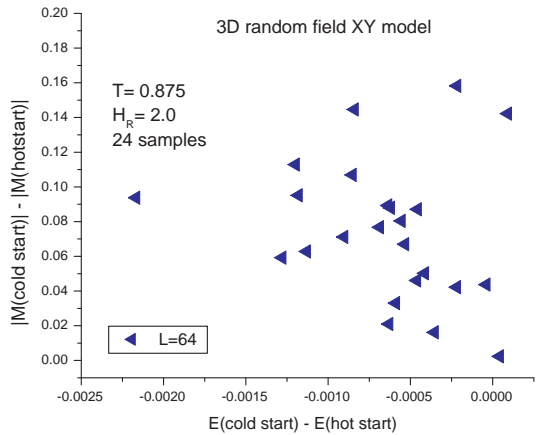}
\caption{\label{Fig.4}(color online) Jump in the magnetization vs.
jump in the energy for simple cubic lattices with $h_r$ = 2.0 at
$T$ = 0.875 with $L$ = 64.  States with hot start and ordered start
initial conditions are compared for each sample.}
\end{figure}

The center of the $L$ = 64 distribution is now at $\delta |M| = 0.075 \pm 0.008$
and $\delta E = -0.00066 \pm 0.00010$.  The average specific heat of the
$L$ = 64 samples at $T$ = 0.875 and $h_r$ = 2.0 is $0.4266 \pm 0.0015$ for
the cooled samples, and $0.4270 \pm 0.0013$ for the heated samples.  We see
again that the specific heat appears to be slightly lower for the cooled
samples than for the heated samples, but the difference is not statistically
significant.  Overall, we see that the difference between the cooled samples
and the heated samples is somewhat smaller on the phase transition line at
$h_r$ = 2.0 than it is for $h_r$ = 1.875.  The width of the $M^2$ distribution,
$T \chi (\vec{\bf k} = 0)$, for these $L$ = 64 samples is $9.08 \pm 0.40$ for
the cooled samples and $9.51 \pm 0.58$ for the heated samples.  This reflects
the fact that the dynamical fluctuations are reduced as the local random fields
become stronger.

Fig.~5 and Fig.~6 show data corresponding to Fig.~2 and Fig.~3, respectively,
at the point $T$ = 0.875 and $h_r$ = 2.0.  Since the $L$ for these figures is
half of the $L$ in Fig.~2 and Fig.~3, the smallest value of $k$ shown is twice
what it was before.  We see that the qualitative behavior does not change when
we move along the phase transition line, but the magnitude of the logarithmic
peak is getting weaker as $h_r$ increases.

The author thinks it is worth observing that the kind of jumps we are seeing in
the energy and the magnetization of finite samples would need to disappear in
the limit $T$ = 0. The  multicritical critical point hypothesis for the behavior
of random field models at $T$ = 0 says that $T$ should be an irrelevant variable
at that point.  However, the behavior we are seeing along the phase transition
line is not consistent with that hypothesis.

\begin{figure}
\includegraphics[width=3.4in]{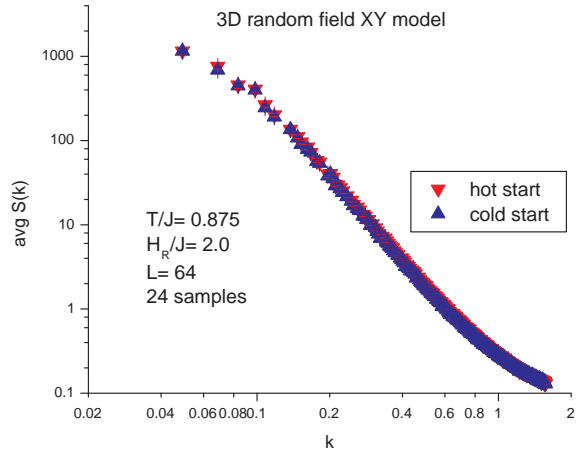}
\caption{\label{Fig.5}(color online) Angle-averaged structure factor
vs. $k$ for $64 \times 64 \times 64$ lattices with $h_r$ = 2.0 at
temperature $T$ = 0.875. Both the $x$-axis and the $y$-axis are scaled
logarithmically.  One $\sigma$ statistical errors are approximately
the size of the plotting symbols.}
\end{figure}

\begin{figure}
\includegraphics[width=3.4in]{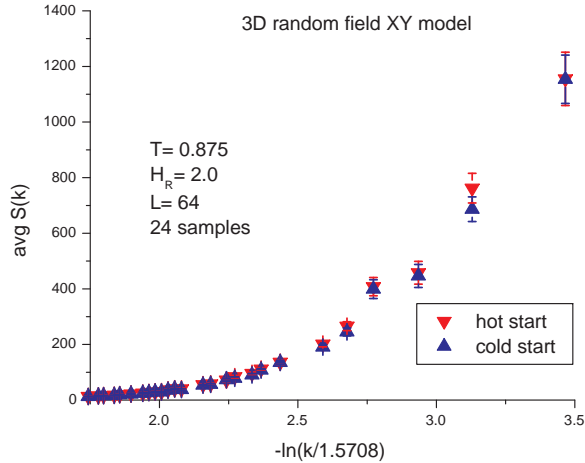}
\caption{\label{Fig.6}(color online) Angle-averaged structure factor
vs. ($-~\ln(k/1.5708)$) for $64 \times 64 \times 64$ lattices with
$h_r$ = 2.0 at $T$ = 0.875.  One $\sigma$ statistical errors
are shown.}
\end{figure}

\section{Discussion}

It is straightforward to calculate the interaction energy of the
spins with the random field.  We merely need to calculate the value
of the second sum in the Hamiltonian as a function of the temperature.
When this is done at $h_r$ = 1.875, it turns out that the value of
the random-field energy has a maximum at about $T$ = 1.75.  Below that
temperature, the ferromagnetic bonds become increasingly successful
in pulling the directions of the local spins away from the directions
of their local random fields.  Of course, there is nothing magic about
$T$ = 1.75.  The temperature at which the maximum value in the
random-field energy will occur will be a function of the value of $h_r$.
This effect is not accounted for in the Imry-Ma argument.

Finding that $S (k)$ diverges at low temperatures in the RFXYM as
$k \to 0$ is not surprising.  This behavior follows from the results of
A. Aharony\cite{Aha78} for models which have a probability distribution
for the random fields which is not isotropic.  According to Aharony's
calculation, if this distribution is even slightly anisotropic, then
we should see a crossover to RFIM behavior.  We know\cite{Imb84,BK87}
that in $d = 3$ the RFIM is ferromagnetic at low temperature if the
random fields are not very strong.  The instability to even a small
anisotropy in the random field distribution should induce a diverging
response in $S (k)$ as $k \to 0$ for the RFXYM in $d$ = 3.  A similar
effect in a related, but somewhat different, model was found by Minchau
and Pelcovits.\cite{MP85}.

More recently, models of quantum-mechanical spins in random fields have
been studied at $T = 0$.\cite{Car13,AN18}  These calculations find
logarithmic divergences of the structure factor as $k \to 0$ in these
quantum versions of random field models.  It is not clear yet that
one should be able to map the classical RFXYM at finite temperature
onto a quantum model at $T$ = 0.  However, A. Aharony's argument about
the instability in the 3D RFXYM makes this connection plausible.

There has been no attempt in this work to equilibrate samples at
temperatures below the apparent phase transition temperature.  Therefore,
we have no data which directly address the question of whether the RFXYM
shows true ferromagnetism in $d$ = 3.  If we assume that the average
$|M|$ of finite samples is subextensive, i.e. the net magnetic moment
grows more slowly than $L^3$ as $L \to \infty$, then it would follow
from the above argument that there might not be any divergence of
$S (k)$ for $k \to 0$ in $d$ = 3 in the cases $n \ge 3$.  If this were
the case, then the behavior of random field $O (n)$ models in $d$ = 3
would be somewhat parallel to the case of the pure $O (n)$ ferromagnets
in $d$ = 2.

Note that it is only $S$ which diverges for the 3D RFXYM.  Unlike the
situation for the Kosterlitz-Thouless transition, $\chi$ does not seem
to have any long-range behavior.  The difference in the behavior of
$S$ and $\chi$ is due to the fact that the local magnetization,
$\vec{\bf M_{i}}$, has a non-zero average value even at high $T$ in a
random field model.  It is very unclear that the behavior we are
seeing can be attributed to topological defects.\cite{GH96}  What is
going on here is that the $Q_{ij}$ terms in Eqn.~4 are canceling against
the $\langle \cos ( \phi_{i} - \phi_{j} ) \rangle$ terms, and giving a
small net result, even at $T_c$.

It appears to the author that what is going on in this model is a
broken ergodicity transition in the phase space, without any change
in the spatial symmetry.  In that sense, it is similar to a spin-glass
phase transition.  However, a random field model does not have the
two-fold Kramers degeneracy of a spin glass.  Therefore the broken
ergodicity occurs in the random field model in a purer form, without
the extra complication of the two-fold symmetry in the phase space.

The reader may be tempted to object that such a phase transition cannot
be described within the usual formalism of equilibrium statistical
mechanics, based on the canonical partition function
\begin{equation}
  Z ( T )~=~ {\bf Tr}_{\{ \phi_i \} } \exp ( - H / T ) \,  ,
\end{equation}
where $H$ is given in Eqn.~1.  We are thinking now about a particular
sample, so the $\theta_i$ variables are fixed.  For a classical system,
the standard formulas based on $Z$ do not have any dependence on dynamics.
That is the point.  The fact that our Monte Carlo calculation sees that
the hot start states and the cold start states we find at $T = 1.00$ are
not the same means that these results cannot be described by $Z ( T )$.
Our calculation is not finding the partition function.  When the
dynamical relaxation time is infinite over a range of $T$, $Z ( T )$ will
not give us the behavior seen in a laboratory experiment.

The idea of the broken ergodicity transition is exactly that we need to
include dynamics in order to understand what is going on.  It is true that
if we ran the Monte Carlo calculation for any finite lattice a very long
time, the results would eventually converge to $Z ( T )$ for that finite
lattice.  However, there is an order of limits issue.  A broken ergodicity
transition, like all thermodynamic phase transitions, only exists in the
limit of an infinite system.  To get correct results in the thermodynamic
limit, we need to take the limit $L \to \infty$ in an appropriate way.  We
should not take the limit of infinite time while holding $L$ fixed.  The
results which come from a Monte Carlo calculation may be thought of as
telling us that the lower critical dimension of the RFXYM is three space
dimensions and one time dimension.  A helpful review of Monte Carlo
calculations, which discusses critical slowing down of the dynamical
behavior at a phase transition, has been given by Sokol.\cite{Sok92}  One
could say that, for the RFXYM problem, critical slowing down is not a bug,
it is a feature.

About six years ago, numerical studies of the RFXYM were performed
by Garanin. Chudnovsky and Proctor.\cite{GCP13}  These authors were
interested in studying lattices of very large $L$.  Such lattices
were much too large for the simulations to be able to reach a thermal
equilibrium, and they did not use any Boltzmann factors in their
dynamics.  Thus the results are some kind of simulated annealing, and
it is not clear what the meaning of their end states is.  The work
being reported here always used Boltzmann factors to relax the state
of the lattice.  It is not possible to make any quantitative comparison,
because they only study low energy states, and give no results for the
behavior at the phase transition.  In further work,\cite{PGC14} these
authors extend their methods to models with other numbers of spin components.
They claim that the 3D $n$ = 3 spin model in a random field of $h_r$ = 1.5
also has a stable ferromagnetic phase at low temperature, but do not
give an estimate of $T_c$.  They also claim that for 2D, the RFXYM
has a ferromagnetic state for $h_r$ = 0.5, which is surely incorrect.
Even the RFIM has no ferromagnetism in 2D.  Therefore the reliability of
their methods is highly questionable.  The functional renormalization group
calculations\cite{TT06,LW06,TT08} do not give any support for the existence
of a ferromagnetic phase for the $n$ = 3 random field case for $d \leq 4$.

There is another model which is more similar to the random-field $XY$
model than the random-field Ising model is.  That model is the 3-state
Potts model in a random field (RFPM).  In the absence of the random-field
term, a 3D 3-state Potts model has a first-order phase transition,
with a substantial latent heat at $T_c$.  In 1989, two groups presented
independent arguments showing that models like this should no longer
have a latent heat when the random-field term is added to the Hamiltonian.
Aizenman and Wehr\cite{AW89} proved that the latent heat must vanish in
the limit $L \to \infty$.  However, the 3D 3-state Potts model presumably
still has a ferromagnetic phase below $T_c$ for weak random fields.  This
author expects that the 3-state Potts model in $d = 3$ will also have a
nonferromagnetic phase of type studied here, for random fields in an
intermediate range of strength.  The Monte Carlo techniques used in this
worked may be able to find such a phase.

Hui and Berker\cite{HB89} argued that the vanishing of the latent heat
implied that a critical fixed point should exist.  This author does not
see, however, why such a fixed point, with its associated divergent
correlation length, should generally exist in a model which has no
translation symmetry, except in those cases where the randomness is an
irrelevant operator.\cite{Har74}  It is certainly true that there are
some cases where such fixed points have been found using $\epsilon$-expansion
calculations.  Subextensive singularities in the specific heat and the
magnetization are completely consistent with the Aizenman-Wehr Theorem.\cite{AW89,AW90}

\section{Summary}

In this work we have performed Monte Carlo studies of the 3D RFXYM on
$L =$ 64, 96 and 128 simple cubic lattices, with a random field strength of
$h_r = 1.875$, and $L = 64$ for $h_r =2.0$.  We compared the properties of
slowly cooled states and slowly heated states at $T = 1.00$ for $h_r = 1.875$,
and $T = 0,875$ for $h_r = 2.0$, which are our estimates of the temperature at
which there appears to be a phase transition.  We display results for the change
in energy and the change in magnetization at this temperature, as a function of
the lattice size.  At the phase transition we measure small jumps in the
magnetization per spin and the energy per spin.  However, it appears that
these jumps are subextensive, meaning that they would scale to zero as
$L \to \infty$.  We estimate that $\delta |M(L)|$ scales like $L^{- 3/4}$, and
$\delta E(L)$ scales like $L^{- 3/2}$.  We have no good reason, however, to
believe that these scaling exponents cannot vary along the phase boundary.
We also compute results for the structure factor, $S ( k )$, and the magnetic
susceptibility, $\chi (\vec{\bf k} = 0)$, under these conditions.  $S ( k )$
appears to be have an approximately logarithmic divergence in the small $k$
limit, but $\chi (\vec{\bf k} = 0)$ seems to have a value of about 14.2 for
$h_r$ = 1.875, at $T$ = 1.00, and is smaller for $h_r$ = 2.0 at $T$ = 0.875.
These characteristics are consistent with the idea that the lower critical
dimension of this model is exactly three.  These results appear to be related
to recent work on quantum disorder.\cite{AN18}

\begin{acknowledgments}
The author thanks N. Sourlas for a helpful conversation about the recent work
on the random field Ising model, and Ofer Aharony for a discussion of his recent
results on quantum disordered models. This work used the Extreme Science and
Engineering Discovery Environment (XSEDE) Bridges Regular Memory at the Pittsburgh
Supercomputer Center through allocations DMR170067 and DMR180003.  The author
thanks the staff of the PSC for their help.

\end{acknowledgments}



\end{document}